# Recurrent neural network-based volumetric fluorescence microscopy

Luzhe Huang[1,2,3], Yilin Luo[1], Yair Rivenson[1,2,3*], Aydogan Ozcan[1,2,3,4*]

[1] *Electrical and Computer Engineering Department, University of California, Los Angeles, California 90095, USA*
[2] *Bioengineering Department, University of California, Los Angeles, California 90095, USA*
[3] *California Nano Systems Institute (CNSI), University of California, Los Angeles, California 90095, USA*
[4] *David Geffen School of Medicine, University of California, Los Angeles, California 90095, USA*
\* *ozcan@ucla.edu, rivensonyair@ucla.edu*

**Abstract:** Volumetric imaging of samples using fluorescence microscopy plays an important role in various fields including physical, medical and life sciences. Here we report a deep learning-based volumetric image inference framework that uses 2D images that are sparsely captured by a standard wide-field fluorescence microscope at arbitrary axial positions within the sample volume. Through a recurrent convolutional neural network, which we term as Recurrent-MZ, 2D fluorescence information from a few axial planes within the sample is explicitly incorporated to digitally reconstruct the sample volume over an extended depth-of-field. Using experiments on *C. Elegans* and nanobead samples, Recurrent-MZ is demonstrated to increase the depth-of-field of a 63 ×/1.4NA objective lens by approximately 50-fold, also providing a 30-fold reduction in the number of axial scans required to image the same sample volume. We further illustrated the generalization of this recurrent network for 3D imaging by showing its resilience to varying imaging conditions, including e.g., different sequences of input images, covering various axial permutations and unknown axial positioning errors. Recurrent-MZ demonstrates the first application of recurrent neural networks in microscopic image reconstruction and provides a flexible and rapid volumetric imaging framework, overcoming the limitations of current 3D scanning microscopy tools.

## 1. Introduction

High-throughput imaging of 3D samples is of significant importance for numerous fields. Volumetric imaging is usually achieved through optical sectioning of samples using various microscopy techniques. Generally, optical sectioning can be categorized based on its dimension of sectioning: (i) 0-dimensional point-wise sectioning, including e.g., confocal [1], two-photon [2] and three-photon [3] laser scanning microscopy, and time-domain optical coherence tomography (TD-OCT) [4]; (ii) 1-dimensional line-wise sectioning, including e.g., spectral domain OCT [5,6] , (iii) 2-dimensional plane-wise sectioning, including e.g., wide-field and light-sheet [7] fluorescence microscopy. In all of these modalities, serial scanning of the sample volume is required, which limits the imaging speed and throughput, reducing the temporal resolution, also introducing potential photobleaching on the sample. Different imaging methods have been proposed to improve the throughput of scanning-based 3D microscopy techniques, such as multifocal imaging [8–13], light-field microscopy [14,15], microscopy with engineered point spread functions (PSFs) [16–18] and compressive sensing [19–21]. Nevertheless, these solutions introduce trade-offs, either by complicating the microscope system design, compromising the image quality and/or resolution or prolonging the image post-processing time. In addition to these, iterative algorithms that aim to solve the inverse 3D imaging problem from a lower dimensional projection of the volumetric image data, such as the fast iterative shrinkage and thresholding algorithm (FISTA) [22] and alternating direction method of multiplier (ADMM) [23] are relatively time-consuming and unstable, and further require user-defined regularization of the optimization process as well as an accurate forward model of the imaging system. Some of these limitations and performance trade-offs have partially restricted the wide-scale applicability of these computational methods for 3D microscopy.

In recent years, emerging deep learning-based approaches have enabled a new set of powerful tools to solve various inverse problems in microscopy [24,25], including e.g., super-resolution imaging [26,27], virtual labeling of specimen [28–32], holographic imaging [33,34], Fourier ptychography microscopy [35], single-shot autofocusing [36,37], three-dimensional image propagation [38], among many others [39]. Benefiting from the recent advances in deep learning, these methods require minimal modification to the underlying microscopy hardware, and result in enhanced imaging performance in comparison to conventional image reconstruction and post-processing algorithms.

The majority of these neural networks applied in microscopic imaging were designed to perform inference using a *single* 2D input image. An alternative method to adapt a deep network's inference ability to utilize information that is encoded over volumetric inputs (instead of a single 2D input image) is to utilize 3D convolution kernels. However, this approach requires a significant number of additional trainable parameters and is therefore more susceptible to overfitting. Moreover, simply applying 3D convolution kernels and representing the input data as a sequence of 2D images would constrain the input sampling grid and introduce practical challenges. As an alternative to 3D convolution kernels, recurrent neural networks (RNNs) were originally designed for sequential temporal inputs, and have been successfully applied in various tasks in computer vision [40-42].

Here, we introduce the first RNN-based volumetric microscopy framework, which is also the first application of RNNs in microscopic image reconstruction; we termed this framework as Recurrent-MZ. Recurrent-MZ permits the digital reconstruction of a sample volume over an extended depth-of-field (DOF) using a few different 2D images of the sample as inputs to a trained RNN (see Fig. 1(a)). The input 2D images are sparsely sampled at arbitrary axial positions within the sample volume and the convolutional recurrent neural network (Recurrent-MZ) takes these 2D microscopy images as its input, along with a set of digital propagation matrices (DPMs) which indicate the *relative distances* ($dz$) to the desired output plane(s). Information from the input images is separately extracted using sequential convolution blocks at different scales, and then the recurrent block aggregates all these features from the previous scans/images, allowing flexibility in terms of the length of the input image sequence as well as the axial positions of these input images, which do not need to be regularly spaced or sampled; in fact, the input 2D images can even be randomly permuted.

We demonstrate the efficacy of the Recurrent-MZ using multiple fluorescent specimens. First, we demonstrate Recurrent-MZ inference for 3D imaging of *C. Elegans* samples, and then quantify its performance using fluorescence nanobeads. Our results demonstrate that Recurrent-MZ increases the depth-of-field of a $63 \times /1.4 NA$ objective lens by approximately 50-fold, also providing a 30-fold reduction in the number of axial scans required to image a sample volume. Furthermore, we demonstrate the robustness of this framework and its inference to axial permutations of the input images as well to uncontrolled errors and noise terms in the axial positioning of different input image scans.

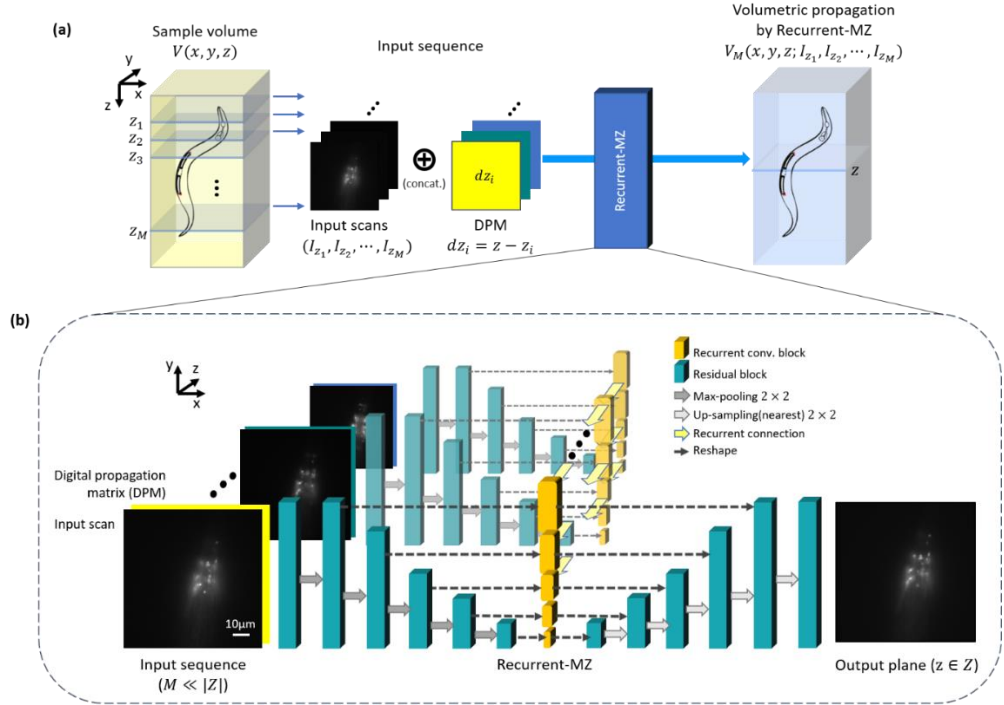

**Fig. 1** Volumetric imaging through Recurrent-MZ. (a) Recurrent-MZ volumetric imaging framework. $M$ is the number of input scans (2D images), and each input scan is paired with its corresponding DPM (Digital Propagation Matrix). (b) Recurrent-MZ network structure. The network and training details are elucidated in the Methods section. $Z$ is the set of all the axial positions within the target sample volume, composed of $|Z|$ unique axial planes. Typically, M=2 or 3 and $|Z| \gg M$.

## 2. Results

We formulate the target sample volume $V(x, y, z)$ as a random field on the set of all axial positions $Z$, i.e., $I_z \in \mathbb{R}^{m \times n}, z \in Z$, where $x, y$ are pixel indices on the lateral plane, $m, n$ are the lateral dimensions of the image, and $z$ is a certain axial position in $Z$. The distribution of such random fields is defined by the 3D distribution of the sample of interest, the PSF of the microscopy system, the aberrations and random noise terms present in the image acquisition system. Recurrent-MZ takes in a set of $M$ 2D axial images, i.e., $\{I_{z_1}, I_{z_2}, \cdots, I_{z_M}\}, 1 < M \ll |Z|$, where $|Z|$ is the cardinality of $Z$, defining the number of unique axial planes in the target sample. The output inference of Recurrent-MZ estimates (i.e., reconstructs) the volume of the sample and will be denoted as $V_M(x, y, z; I_{z_1}, I_{z_2}, \cdots, I_{z_M})$. Starting with the next sub-section we summarize Recurrent-MZ inference results using different fluorescent samples.

### 2.1 Recurrent-MZ based volumetric imaging of C. Elegans samples

A Recurrent-MZ network was trained and validated using *C. Elegans* samples, and then blindly tested on new specimens that were *not* part of the training/validation dataset. This trained Recurrent-MZ was used to reconstruct *C. Elegans* samples with high fidelity over an extended axial range of 18 $\mu m$ based on three 2D input images that were captured with an axial spacing of $\Delta z = 6\mu m$; these three 2D images were fed into Recurrent-MZ in groups of two, i.e., M=2 (Fig. 2). The comparison images of the same sample volume were obtained by scanning a wide-field fluorescence microscope with a $63 \times /1.4NA$ objective lens and capturing $|Z| = 91$ images with an axial spacing of $\Delta z = 0.2\mu m$ (see the Methods section). The inference performance of Recurrent-MZ is both qualitatively and quantitatively demonstrated in Fig. 2 and Video S1. Even in the middle of two adjacent input images (see the $z = 11.4\ \mu m$ row of Fig. 2), Recurrent-MZ is able to output images with a very good match to the ground truth image, achieving a normalized root mean square error (NRMSE) of 6.45 and a peak signal-to-noise ratio (PSNR) of 33.96. Moreover, when the output axial position is selected to be adjacent to one of the input images (see the $z = 1.6\ \mu m$ row of Fig. 2), the Recurrent-MZ even outperforms the nearest neighbour interpolation, achieving NRMSE=2.33 and PNSR=36.47. As also highlighted in Video S1, Recurrent-MZ is able to significantly extend the axial range of the reconstructed images using only three 2D input scans, each captured with a $1.4NA$ objective lens that has a depth-of-field of $0.4\mu m$.

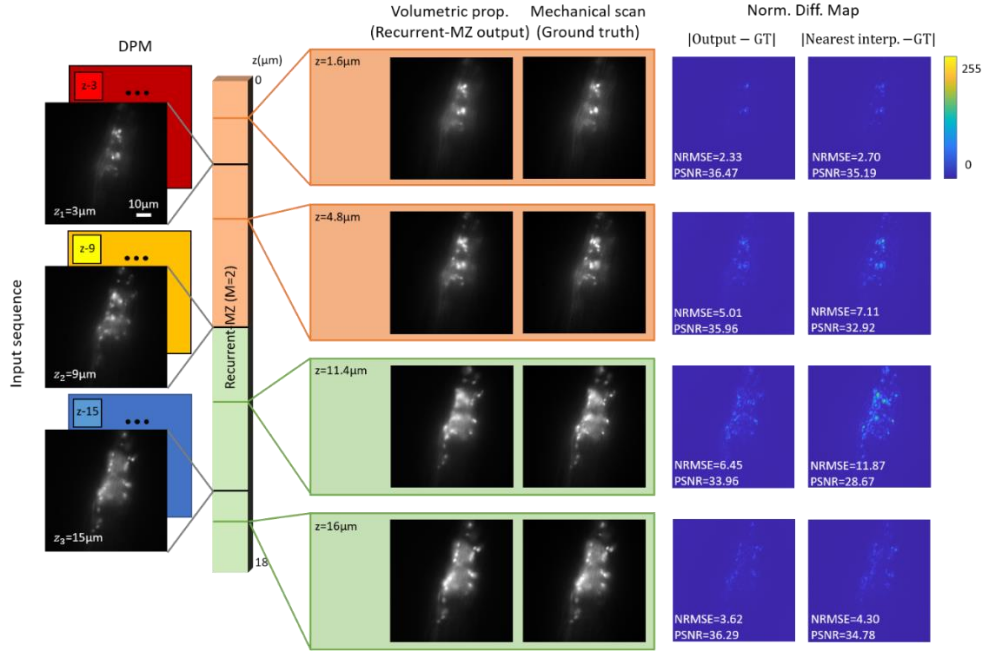

**Fig. 2** Volumetric imaging of *C. Elegans* from sparse wide-field scans using Recurrent-MZ. The DPMs in the input sequence are used to define an arbitrary axial position (z) within the sample volume. In this implementation, Recurrent-MZ takes in 2 input scans (*M*=2) to infer the image of an output plane, as indicated by the color of each output box. See Video S1 to compare the reconstructed sample volume inferred by Recurrent-MZ against the ground truth, |Z|=91 images captured with an axial step size of 0.2μm.

It is worth noting that although Recurrent-MZ presented in Fig. 2 was trained with 2 input images (i.e., *M*=2), it still can be fed with $M \geq 3$ input images thanks to its recurrent scheme. *Regardless of the choice of M , all Recurrent-MZ networks have the same number of parameters, where the only difference is the additional time that is required during the training and inference phases*; for example the inference time of Recurrent-MZ with $M = 2$ and $M = 3$ for a single output plane ($1024 \times 1024$ pixels) is $0.18\ s$ and $0.28\ s$, respectively. In practice, using a larger *M* yields a better performance in terms of the reconstruction fidelity (see e.g., Fig. S1(a)), at the cost of a trade-off of imaging throughput and computation time. The detailed discussion about this trade-off is provided in the Discussion section.

### 2.2 Recurrent-MZ based volumetric imaging of fluorescence nanobeads

Next, we demonstrated the performance of Recurrent-MZ using $50\ nm$ fluorescence nanobeads. These nanobead samples were imaged through the TxRed channel using a 63 × /1.4NA objective lens (see the Methods section). The Recurrent-MZ model was trained on a dataset with $M = 3$ input images, where the axial spacing between the adjacent planes was $\Delta z = 3\mu m$. The ground truth images of the sample volume were captured by mechanical scanning over an axial range of $10\mu m$, i.e., $|Z| = 101\ images\ with\ \Delta z = 0.1\mu m$ were obtained. Figure 3 shows both the side views and the cross-sections of the sample volume reconstructed by Recurrent-MZ (M=3), compared against the $|Z| = 101$ images captured through the mechanical scanning of the same sample. The first column of Fig. 3(a) presents the M=3 input images and their corresponding axial positions, which are also indicated by the blue dashed lines. Through the quantitative histogram comparison shown in Fig. 3(b), we see that the reconstructed volume by Recurrent-MZ matches the ground truth volume with high fidelity. For example, the full width at half maximum (FWHM) distribution of individual nanobeads inferred by Recurrent-MZ (mean FWHM = $0.4401\mu m$) matches the results of the ground truth (mean FWHM = $0.4428\mu m$) very well. We also showed the similarity of the ground truth histogram with that of the Recurrent-MZ output by calculating the Kullback-Leibler (KL) divergence, which is a distance measure between two distributions; the resulting KL divergence of 1.3373 further validates the high fidelity of Recurrent-MZ reconstruction when compared to the ground truth, acquired through $|Z| = 101$ images captured via mechanical scanning of the sample with $\Delta z = 0.1\mu m$.

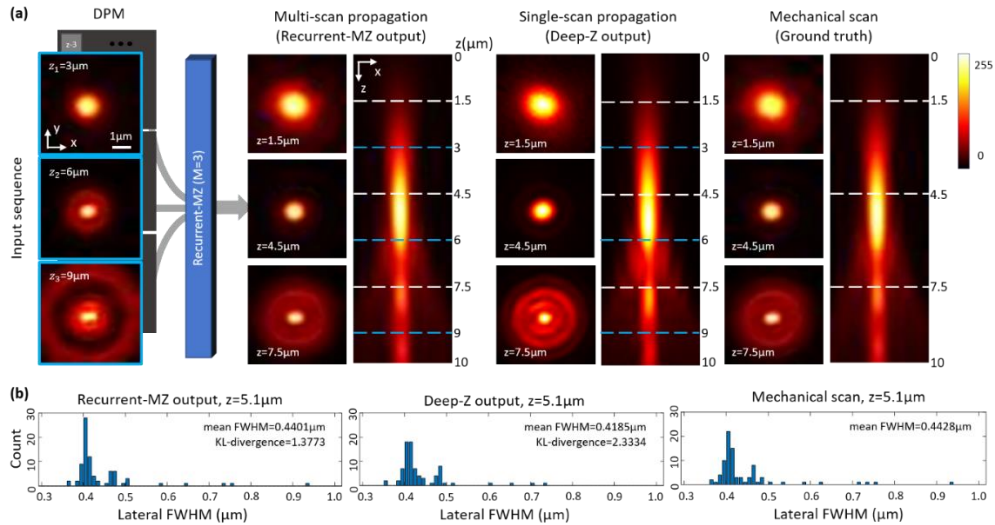

**Fig. 3** The performance of Recurrent-MZ using fluorescence nanobeads. (a) Volumetric imaging using Recurrent-MZ (M=3) and Deep-Z on 50nm fluorescence nanobeads. There are 3 input images for Recurrent-MZ (M=3) and to provide a fair comparison, Deep-Z always takes in the nearest input image among these 3 inputs to infer another axial plane. The PSFs generated by Recurrent-MZ, Deep-Z and mechanical scanning ($\Delta z = 0.1\mu m$) are shown for comparison. (b) Lateral FWHM histograms for 88 individual isolated fluorescence nanobeads, measured from mechanical scanning (101 axial images), Deep-Z reconstruction and Recurrent-MZ reconstruction ($M = 3$). Also see Video S2.

Figure 3 also reports the comparison of Recurrent-MZ inference results with respect to another fluorescence image propagation network termed Deep-Z [38]. Deep-Z is designed for taking a *single* 2D image as input, and therefore there is an inherent trade-off between the propagation quality and the axial refocusing range (from a given focal plane), which ultimately limits the effective volumetric space-bandwidth-product (SBP) that can be achieved using Deep-Z. In this comparison between Recurrent-MZ and Deep-Z (Fig. 3), the nearest input image is used for Deep-Z based propagation; in other words, three non-overlapping volumes are separately inferred using Deep-Z from the input scans at $z = 3, 6$ and $9\mu m$, respectively (this provides a fair comparison against Recurrent-MZ with M=3 input images). As illustrated in Fig. 3(b), Deep-Z inference resulted in a mean FWHM of $0.4185\mu m$ and a KL divergence of 2.3334, which illustrate the inferiority of single-image-based volumetric propagation, when compared to the results of Recurrent-MZ. The same conclusion regarding the performance comparison of Recurrent-MZ and Deep-Z inference is further supported using the *C. Elegans* imaging data reported in Fig. 2 (Recurrent-MZ) and in Fig. S2 (Deep-Z). For example, Deep-Z inference results in an NRMSE of 8.02 and a PSNR of 32.08, while Recurrent-MZ (M=2) improves the inference accuracy, achieving an NRMSE of 6.45 and a PSNR of 33.96.

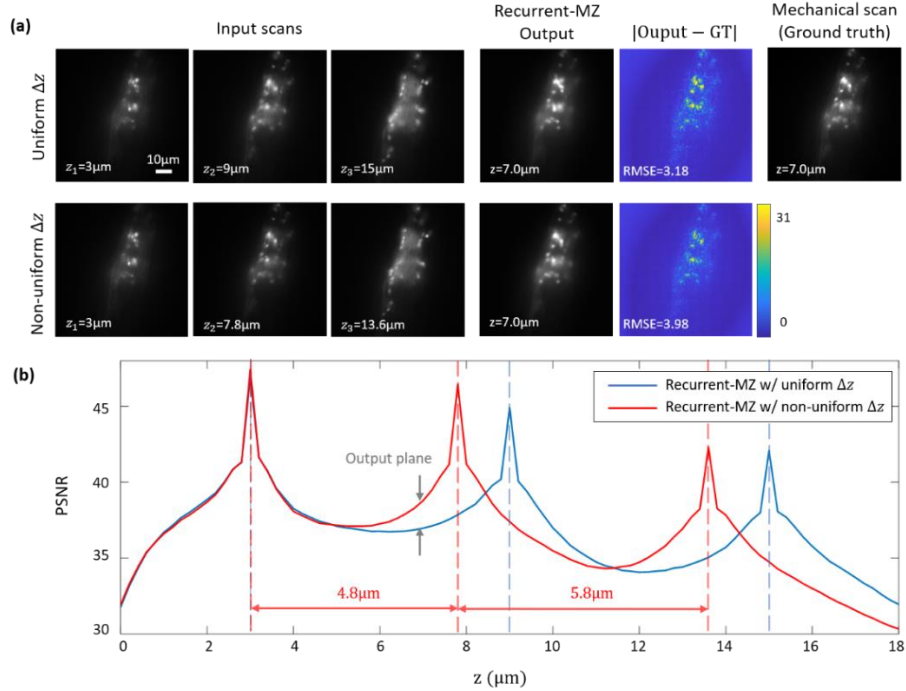

**Fig. 4** Generalization of Recurrent-MZ to non-uniformly spaced input images. (a) Recurrent-MZ was trained on *C. Elegans* samples with equidistant inputs (M=3, $\Delta z = 6 \mu m$), and blindly tested on both uniformly sampled and non-uniformly sampled input images of new samples. (b) The PSNR values of the output images (with uniformly spaced and non-uniformly spaced input images) are calculated with respect to the ground truth, corresponding image. Blue: Outputs of Recurrent-MZ (M=3) for uniformly spaced inputs, Red: Outputs of Recurrent-MZ (M=3) for non-uniformly spaced inputs. Dashed lines indicate the axial positions of the input 2D images.

*2.3 Generalization of Recurrent-MZ to non-uniformly sampled input images*

Next, we demonstrated, through a series of experiments, the generalization performance of Recurrent-MZ on ***non-uniformly*** sampled input images, in contrast to the training regiment, which only included uniformly spaced inputs. These non-uniformly spaced input image planes were randomly selected from the same testing volume as shown in Fig. 2, with the distance between two adjacent input planes made smaller than the uniform axial spacing used in the training dataset ($\Delta z = 6\mu m$). Although the Recurrent-MZ was solely trained with *equidistant* input scans, it generalized to successfully perform volumetric image propagation using *non-uniformly sampled* input images. For example, as shown in Fig. 4(a), the input images of Recurrent-MZ were randomly selected at $(z_1, z_2, z_3) = (3, 7.8, 13.6)\ \mu m$, respectively, and the output inference at $z = 7.0\ \mu m$ very well matches the output of Recurrent-MZ that used uniformly sampled inputs acquired at $(z_1, z_2, z_3) = (3, 9, 15)\ \mu m$, respectively. Figure 4(b) further demonstrates the inference performance of Recurrent-MZ using non-uniformly sampled inputs throughout the specimen volume. The blue (uniform inputs) and the red curves (non-uniform inputs) in Fig. 4(b) have very similar trends, illustrating the generalization of Recurrent-MZ, despite being only trained with uniformly-sampled input images with a fixed $\Delta z$.

We further investigated the effect of the hyperparameter $\Delta z$ on the performance of Recurrent-MZ. For this, three different Recurrent-MZ networks were trained using $\Delta z = 4, 6,$ and $8\mu m$, respectively, and then blindly tested on a new input sequence with $\Delta z = 6\ \mu m$. Fig. S3 shows the trade-off between the peak performance and the performance consistency over the inference axial range: by decreasing $\Delta z$, Recurrent-MZ demonstrates a better peak inference performance, indicating that more accurate propagation has been learned from smaller $\Delta z$, whereas the variance of PSNR, corresponding to the performance consistency over a larger axial range, is degraded for smaller $\Delta z$.

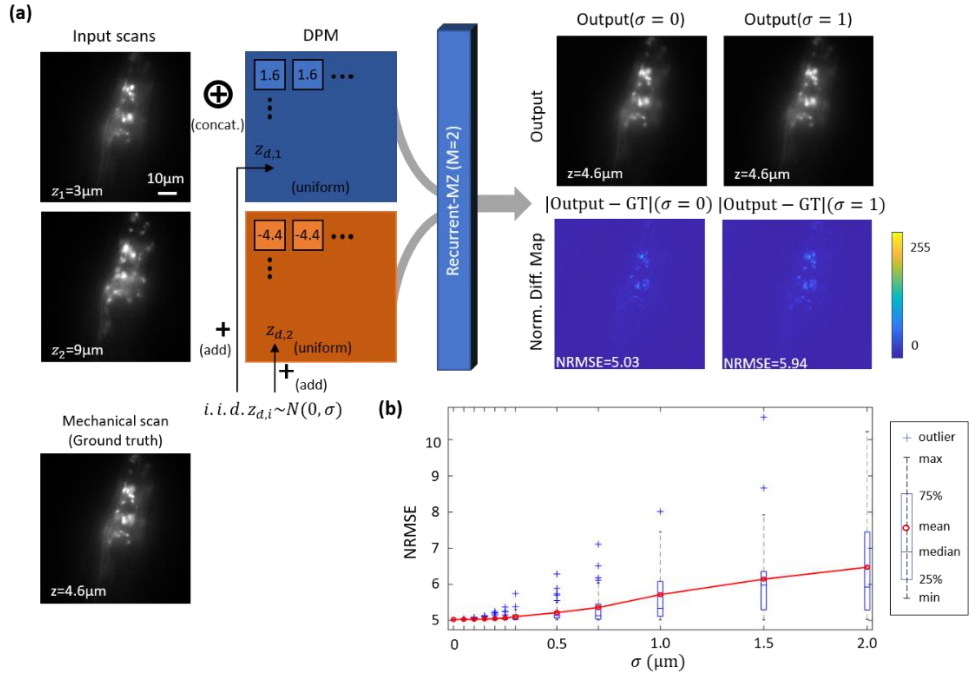

**Fig. 5** Stability test of Recurrent-MZ inference. (a) An additive Gaussian noise with zero mean and a standard variance of $\sigma$ was injected into each DPM to test the stability of Recurrent-MZ inference. The output images and difference maps (with respect to ground truth) with no injected noise ($\sigma = 0$) and $\sigma = 1\mu m$ are shown. (b) The NRMSE – $\sigma$ boxplot. NRMSE values were calculated over 50 random tests. The difference maps were normalized by the maximum difference between the input images and the ground truth.

### 2.4 Inference stability of Recurrent-MZ

During the acquisition of the input scans, inevitable measurement errors are introduced by e.g., PSF distortions and focus drift [43], which jeopardize both the precision and accuracy of the axial positioning measurements. Hence, it is necessary to take these effects into consideration and examine the stability of the Recurrent-MZ inference. For this, Recurrent-MZ was tested on the same image test set as in Figure 2, only this time, independent and identically distributed (i.i.d.) Gaussian noise was injected into the DPM of each input image, mimicking the measurement uncertainty when acquiring the axial scans. The noise was added to the DPM as follows:

$$Z_{i,noised} = Z_i + z_{d,i} J, \quad i = 1, 2, \cdots, M,$$

where $Z_i$ is the DPM ($m \times n$ matrix) of the i-th input image, $z_{d,i} \sim N(0, \sigma^2), i = 1, 2, \cdots, M$ and $J$ is an all-one $m \times n$ matrix.

The results of this noise analysis reveals that, as illustrated in Fig. 5(b), the output images of Recurrent-MZ ($M = 2$) degrade as the variance of the injected noise increases, as expected. However, even at a relatively significant noise level, where the microscope stage or sample drift is represented with a standard variation of $\sigma = 1 \mu m$ (i.e., 2.5-fold of the objective lens depth-of-field, 0.4 $\mu m$), Recurrent-MZ inference successfully matches the ground truth with an NRMSE of 5.94; for comparison, the baseline inference (with $\sigma = 0 \mu m$) has an NRMSE of 5.03, which highlights the resilience of Recurrent-MZ framework against axial scanning errors and/or uncontrolled drifts in the sample/stage.

*2.5 Permutation invariance of Recurrent-MZ*

Next, we focused on *post hoc* interpretation [44,45] of the Recurrent-MZ framework, *without* any modifications to its design or the training process. For this, we explored to see if Recurrent-MZ framework exhibits *permutation invariance*, i.e.,

$$V_M(I_1, I_2, \cdots, I_M) = V_M(I_{i_1}, I_{i_2}, \cdots, I_{i_M}), \quad \forall (i_1, i_2, \cdots, i_M) \in S_M,$$

where $S_M$ is the permutation group of $M$. To explore the permutation invariance of Recurrent-MZ (see Fig. 6), the test set's input images were randomly permuted, and fed into the Recurrent-MZ (M=3), which was *solely trained with input images sorted by z*. We then quantified Recurrent-MZ outputs over all the 6 permutations of the M=3 input images, using the average RMSE ($\mu_{RMSE}$) and the standard variance of the RMSE ($\sigma_{RMSE}$), calculated with respect to the ground truth image $I$:

$$\mu_{RMSE} = \frac{1}{6} \sum_{(i_1, i_2, i_3) \in S_3} \text{RMSE}(V_{iii}(I_{i_1}, I_{i_2}, I_{i_3}), I),$$

$$\sigma_{RMSE} = \sqrt{\frac{1}{6} \sum_{(i_1, i_2, i_3) \in S_3} \left(\text{RMSE}(V_{iii}(I_{i_1}, I_{i_2}, I_{i_3}), I) - \mu_{RMSE}\right)^2},$$

where $RMSE(I, J)$ gives the RMSE between image $I$ and $J$. In Fig. 6(e), the red line indicates the average RMSE over 6 permutations and the pink shaded region indicates the standard deviation of RMSE over these 6 permutations. Compared with the blue line in Fig. 6(e), which corresponds to the output of the Recurrent-MZ with the inputs sorted by $z$, the input image permutation results highlight the success of Recurrent-MZ with different input image sequences, despite being trained solely by depth sorted inputs. In contrast, *non-recurrent* convolution neural network (CNN) architectures, such as 3D U-Net [46], inevitably lead to input permutation instability as they require a fixed length and sorted input sequences; this failure of *non-recurrent* CNN architectures is illustrated in Fig. S4.

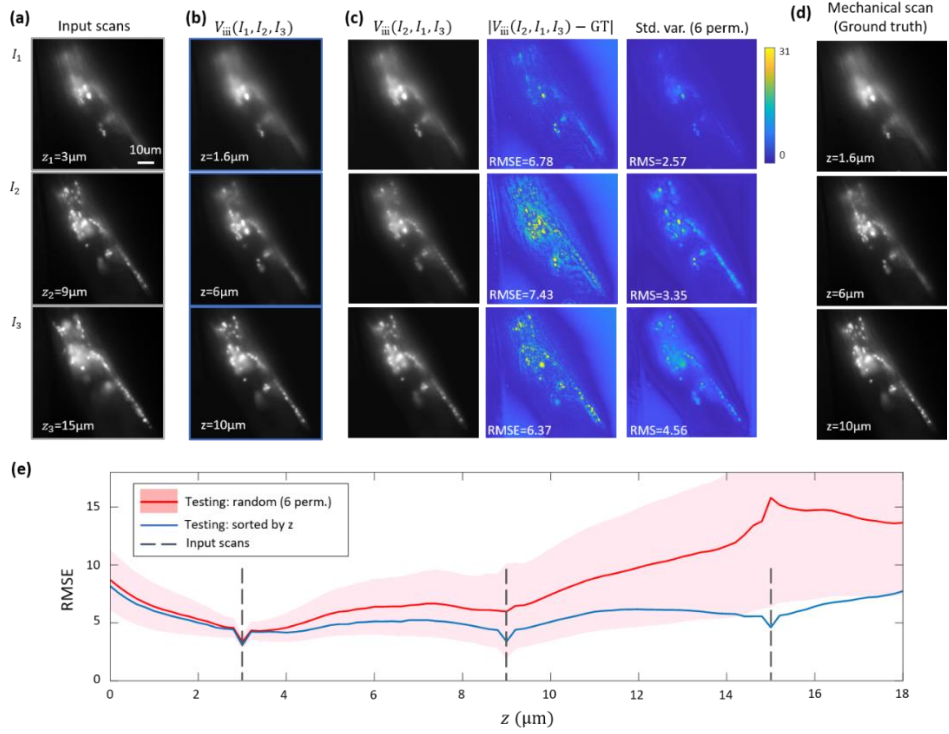

**Fig. 6** Permutation invariance of Recurrent-MZ to the input images. Recurrent-MZ was trained with inputs (M=3) sorted by $z$ and tested on new samples with both inputs sorted by $z$ as well as 6 random permutations of the same inputs to test its permutation invariance. (a) The input scans sorted by $z$, (b) the Recurrent-MZ outputs and the corresponding difference maps of the input sequence $(I_1, I_2, I_3)$, (c) the test output with input sequence $(I_2, I_1, I_3)$, and the pixel-wise standard variance over all the 6 random permutations, (d) the ground truth images obtained by mechanical scanning through the same sample, acquired with an axial spacing of 0.2μm, (e) red solid line: the average RMSE of the outputs of randomly permuted input images; pink shadow: the standard deviation RMSE of the outputs of randomly permuted input images; blue solid line: the RMSE of the output of input images sorted by $z$. The range of grayscale images is 255 while that of the standard variance images is 31.

We also explored different training schemes to further improve the permutation invariance of Recurrent-MZ, including training with input images sorted in descending order by the relative distance (dz) to the output plane as well as randomly sorted input images. As shown in Fig. S5, the Recurrent-MZ trained with input images that are sorted by depth, $z$, achieves the best inference performance, indicated by an NRMSE of 4.03, whereas incorporating randomly ordered inputs in the training phase results in the best generalization for different input image permutations. The analyses reported in Fig. S5 further highlight the impact of different training schemes on the inference quality and the permutation invariance feature of the resulting trained Recurrent-MZ network.

*2.6 Repetition invariance of Recurrent-MZ*

Next, we explored to see if Recurrent-MZ framework exhibits repetition invariance. Figure 7 demonstrates the repetition invariance of Recurrent-MZ when it was repeatedly fed with input image $I_1$. The output images of Recurrent-MZ in Fig. 7(b) show its consistency for 2, 4 and 6 repetitions of $I_1$, i.e., $V_{ii}(I_1, I_1)$, $V_{ii}(I_1, I_1, I_1, I_1)$ and $V_{ii}(I_1, I_1, I_1, I_1, I_1, I_1)$, which resulted in an RMSE of 12.30, 11.26 and 11.73, respectively. Although Recurrent-MZ was never trained with repeated input images, its recurrent scheme still demonstrates the correct propagation under repeated inputs of the same 2D plane. When compared with the output of Deep-Z (i.e., Deep-Z($I_1$)) shown in Fig. 7(c), Recurrent-MZ, with a single input image or its repetitions, exhibits comparable reconstruction quality. Fig. S6 also presents a similar comparison when $M = 3$, further supporting the same conclusion.

While for a single input image ($I_1$ or its repeats) the blind inference performance of Recurrent-MZ is on par with Deep-Z($I_1$), the incorporation of multiple input planes gives a superior performance to Recurrent-MZ over Deep-Z. As shown in the last two columns of Fig. 7(b), by adding another depth image, $I_2$, the output of Recurrent-MZ is significantly improved, where the RMSE decreased to 8.78; this represents a better inference performance compared to Deep-Z($I_1$) and Deep-Z($I_2$) as well as the average of these two Deep-Z outputs (see Figs. 7(b-c)). The same conclusion is further supported in Figs. S6(b-c) for M=3, demonstrating that Recurrent-MZ is able to outperform Deep-Z even if all of its M input images are individually processed by Deep-Z and averaged, showing the superiority of the presented recurrent inference framework.

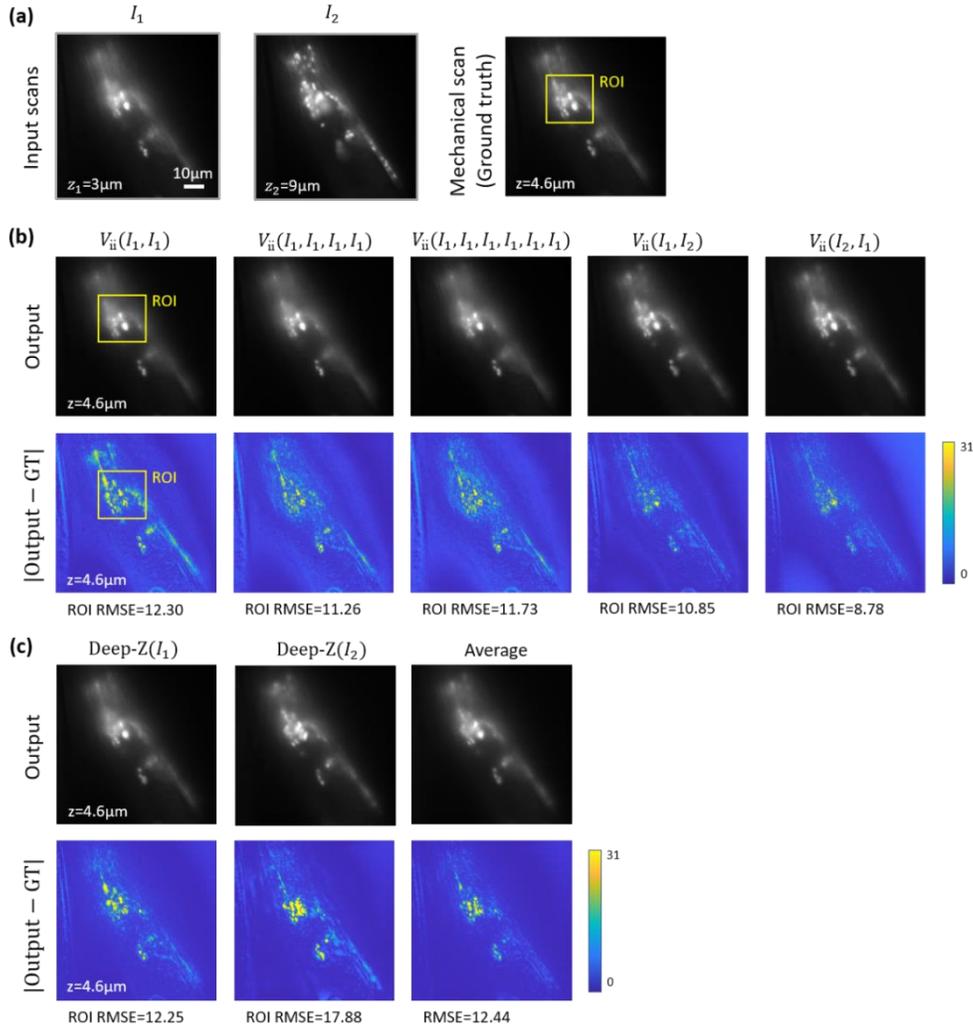

**Fig. 7** Repetition invariance of Recurrent-MZ. Recurrent-MZ was trained with inputs (M=2) sorted by their relative distances (dz) to the output plane, but tested on a new sample by repeatedly feeding the input image ($I_1$) to test its repetition invariance. (a) The input images and the ground truth image obtained by mechanical scanning (with an axial spacing of 0.2μm), (b)

the Recurrent-MZ outputs and the corresponding difference maps of repeated $I_1$, i.e., $V_{ii}(I_1, I_1)$, $V_{ii}(I_1, I_1, I_1, I_1)$ and $V_{ii}(I_1, I_1, I_1, I_1, I_1, I_1)$ as well as $V_{ii}(I_1, I_2)$ and $V_{ii}(I_2, I_1)$, (c) the outputs and corresponding difference maps of Deep-Z with a single input image ($I_1$ or $I_2$), and the pixel-wise average of Deep-Z($I_1$) and Deep-Z($I_2$). All RMSE values are calculated based on the region of interest (ROI) marked by the yellow box. The range of grayscale images is 255 while that of the standard variance images is 31.

## 3. Discussion

We demonstrated a new deep learning-based volumetric imaging framework termed Recurrent-MZ enabled by a convolutional recurrent neural network, which extends the DOF of the microscopy system by around 50-fold from sparse 2D scanning, providing a 30-fold reduction in the number of required mechanical scans. Another advantage of Recurrent-MZ is that it does not require special optical components in the microscopy set-up or an optimized scanning strategy. Despite being trained with equidistant input scans, Recurrent-MZ successfully generalized to use input images acquired with a non-uniform axial spacing as well as unknown axial positioning errors, all of which demonstrate its robustness.

In a practical application, the users of Recurrent-MZ should select an optimum $M$ to provide a balance between the inference image quality of the reconstructed sample volume and the imaging throughput. For example, it is possible to set a stopping threshold, $\epsilon$, for the volumetric reconstruction improvement that is provided by adding another image/scan to Recurrent-MZ, in terms of the Euclidean distance from the volume which was reconstructed from previous images; stated differently, the scanning can stop when e.g., $\|V_M(I_1, \cdots, I_M) - V_{M-1}(I_1, \cdots, I_{M-1})\|_F \leq \epsilon$, where $\|\cdot\|_F$ defines the Frobenius norm.

Importantly, this study shows the first application of convolutional recurrent neural networks in microscopic image reconstruction, and also reveals the potential of RNNs in microscopic imaging when sequential image data are acquired. With regards to solving inverse problems in microscopic imaging, most existing deep learning-based methods are optimized for a single shot/image, whereas sequential shots are generally convenient to obtain and substantial sample information hides in their 3D distribution. In this work, through the incorporation of sequential 2D scans, the presented Recurrent-MZ integrates the information of different input images from different depths to gain considerable improvement in the volumetric image quality and the output DOF. In contrast to 3D convolutional neural networks that generally require a fixed sampling grid, the presented recurrent scheme is compatible with input sequences of variable lengths, as shown in Fig. 7. Another interesting property that we demonstrated is the robustness of Recurrent-MZ in terms of its resilience to input image permutations (Fig. 6), which could lead to catastrophic failure modes for standard convolutional networks, as also illustrated in Fig. S4.

In summary, Recurrent-MZ provides a rapid and flexible volumetric imaging framework with reduced number of axial scans, and opens up new opportunities in machine learning-based 3D microscopic imaging. The presented recurrent neural network structure could also be widely applicable to process sequential data resulting from various other 3D imaging modalities such as OCT, Fourier ptychographic microscopy, holography, structured illumination microscopy, among others.

## 4. Methods

### 4.1 Sample preparation, image acquisition and dataset preparation

The *C. Elegans* samples were firstly cultured and stained with GFP using the strain AML18. AML18 carries the genotype wtfIs3 [rab-3p::NLS::GFP+rab-3p::NLS::tagRFP] and expresses GFP and tagRFP in the nuclei of all the neurons. C. Elegans samples were cultured on nematode growth medium seeded with OP50 E. Coli bacteria using standard conditions. During the imaging process, the samples were washed off the plates with M9 solution and anesthetized with 3 mM levamisole, and then mounted on slides seeded with 3% agarose.

The images of *C. Elegans* were captured by an inverted scanning microscope (TCS SP8, Leica Microsystems), using a $63 \times /1.4NA$ objective lens (HC PL APO $63 \times /1.4NA$ oil CS2, Leica Microsystems) and a FITC filter set (excitation / emission wavelengths: $495nm$ / $519nm$), resulting in a DoF about $0.4\ \mu m$. A monochrome scientific CMOS camera (Leica DFC9000GTC-VSC08298) was used for imaging where each image has $1024 \times 1024$ pixels and 12-bit dynamic range. For each FOV, 91 images with $0.2\mu m$ axial spacing were recorded. A total of 100 FOVs were captured and exclusively divided into training, validation and testing datasets at the ratio of 41:8:1, respectively, where the testing dataset was strictly captured on distinct worms that were not used in training dataset.

The nanobead image dataset consists of wide-field microscopic images that were captured using $50nm$ fluorescence beads with a Texas Red filter set (excitation / emission wavelengths: $589nm$ / $615nm$). The wide-field microscopy system consists of an inverted scanning microscope (TCS SP8, Leica Microsystems) and a $63 \times /1.4NA$ objective lens (HC PL APO $63 \times /1.4NA$ oil CS2, Leica Microsystems). Each volume contains 101 images with $0.1\mu m$ axial spacing. A subset of 400, 86 and 16 volumes were exclusively divided as training, validation and testing datasets.

Each captured image volume was first axially aligned using the ImageJ plugin 'StackReg' [47] for correcting the lateral stage shift and stage rotation. Secondly, an image with extended depth of field (EDF) was generated for each volume, using the ImageJ plugin 'Extended Depth of Field' [48]. The EDF image was later used as a reference for the following image processing steps: (1) apply triangle thresholding to the EDF image to separate the background and foreground contents [38], (2) draw the mean intensity from the background pixels as the shift factor, and the 99% percentile of the foreground pixels as the scale factor, (3) normalize the volume by the shift and scale factors. Thirdly, training FOVs were cropped into small regions of $256 \times 256$ pixels without any overlap. Eventually, the data loader randomly selects $M$ images from the volume with an axial spacing of $\Delta z = 6\mu m$ (*C. Elegans*) and $\Delta z = 3\mu m$ (nanobeads) in both the training and testing phases.

*4.2 Network structure*

Recurrent-MZ is based on a convolutional recurrent network [49] design, which combines the advantages of both convolutional neural networks [39] and recurrent neural networks in processing sequential inputs [50,51]. A common design of the network is formed by an encoder-decoder structure [52,53], with the convolutional recurrent units applying to the latent domain [40,54–56]. Furthermore, inspired by the success of exploiting multiscale features in image translation tasks [57–59], a sequence of cascaded encoder-decoder pairs is utilized to exploit and incorporate image features at different scales from different axial positions.

As shown in Fig. 1(b), the output of last encoder block $x_{k-1}$ is pooled and then fed into the k-th block, which can be expressed as

$$x_k = \text{ReLU}\left(\text{BN}\left(\text{Conv}_{k,2}\left(\text{ReLU}\left(\text{BN}\left(\text{Conv}_{k,1}(\text{P}(x_{k-1}))\right)\right)\right)\right)\right) \qquad (1)$$

where $P(\cdot)$ is the $2 \times 2$ max-pooling operation, $\text{BN}(\cdot)$ is batch normalization, $\text{ReLU}(\cdot)$ is the rectified linear unit activation function and $\text{Conv}_{k,i}(\cdot)$ stands for the i-th convolution layer in the k-th encoder block. The convolution layers in all convolution blocks have a kernel size of $3 \times 3$, with a stride of 1, and the number of channels for $Conv_{k,1}$ and $Conv_{k,2}$ are $20 \cdot 2^{k-2}$ and $20 \cdot 2^{k-1}$, respectively. Then, $x_k$ is sent to the recurrent block, where features from the sequential input images are recurrently integrated:

$$s_k = x_k + \text{Conv}_{k,3}(\text{RConv}_k(x_k)) \qquad (2)$$

where $\text{RConv}_k(\cdot)$ is the convolutional recurrent layer with kernels of $3 \times 3$ and a stride of 1, the $\text{Conv}_{k,3}(\cdot)$ is a $1 \times 1$ convolution layer. Finally, at the decoder part, $s_k$ is concatenated with the up-sampled output from last decoder convolution block, and fed into the k-th decoder block, so the output of k-th decoder block can be expressed as

$$y_k = \text{ReLU}\left(\text{BN}\left(\text{Conv}_{k,5}\left(\text{ReLU}\left(\text{BN}\left(\text{Conv}_{k,4}(\text{I}(y_{k-1}) \oplus s_k)\right)\right)\right)\right)\right) \quad (3)$$

where $\oplus$ is the concatenation operation, $I(\cdot)$ is the $2 \times 2$ up-sampling operation using nearest interpolation and $\text{Conv}_{k,i}(\cdot)$ are the convolution layers of the k-th decoder block.

In this work, the gated recurrent unit (GRU) [60] is used as the recurrent unit, i.e., the $\text{RConv}(\cdot)$ layer in Eq. (2) updates $h_t$, given the input $x_t$, through the following three steps:

$$f_t = \sigma(W_f * x_t + U_f * h_{t-1} + b_f) \quad (4)$$

$$\widehat{h}_t = \tanh(W_h * x_t + U_h * (f_t \odot h_{t-1}) + b_h) \quad (5)$$

$$h_t = (1 - f_t) \odot h_{t-1} + f_t \odot \widehat{h}_t \quad (6)$$

where $f_t, h_t$ are forget and output vectors at time step $t$, respectively, $W_f, W_h, U_f, U_h$ are the corresponding convolution kernels, $b_f, b_h$ are the corresponding biases, $\sigma$ is the sigmoid activation function, $*$ is the 2D convolution operation, and $\odot$ is the element-wise multiplication. Compared with long short term memory (LSTM) network [61], GRU entails fewer parameters but is able to achieve similar performance.

*4.3 Recurrent-MZ implementation*

The Recurrent-MZ was written and implemented using TensorFlow 2.0. In both training and testing phases, a DPM is automatically concatenated with the input image by the data loader, indicating the relative axial position of the input plane to the desired output plane, i.e., the input in the training phase has dimensions of $M \times 256 \times 256 \times 2$. Through varying the DPMs, Recurrent-MZ learns to digitally propagate inputs to any designated plane, and thus forming an output volume with dimensions of $|Z| \times 256 \times 256$.

The training loss of Recurrent-MZ is composed of three parts: (i) pixel-wise BerHu loss [62,63], (ii) multiscale structural similarity index (MSSSIM) [64], and (iii) the adversarial loss using the generative adversarial network (GAN) [65] structure. Based on these, the total loss of Recurrent-MZ, i.e., $L_V$, is expressed as

$$L_V = \alpha \text{BerHu}(\widehat{y}, y) + \beta \text{MSSSIM}(\widehat{y}, y) + \gamma[D(\widehat{y}) - 1]^2 \quad (7)$$

$\widehat{y}$ is the output image of the Recurrent-MZ, and $y$ is the ground truth image for a given axial plane. $\alpha, \beta, \gamma$ are the hyperparameters, which were set as 3, 1 and 0.5, respectively. And the MSSSIM and BerHu losses are expressed as:

$$\text{MSSSIM}(x, y) = \left[\frac{2\mu_{x_M}\mu_{y_M} + C_1}{\mu_{x_M}^2 + \mu_{y_M}^2 + C_1}\right]^{\alpha_M} \cdot \prod_{j=1}^{M} \left[\frac{2\sigma_{x_j}\sigma_{y_j} + C_2}{\sigma_{x_j}^2 + \sigma_{y_j}^2 + C_2}\right]^{\beta_j} \left[\frac{\sigma_{x_j y_j}^2 + C_3}{\sigma_{x_j}\sigma_{y_j} + C_3}\right]^{\gamma_j} \quad (8)$$

$$\text{BerHu}(x, y) = \sum_{\substack{m,n \\ |x(m,n)-y(m,n)| \leq c}} |x(m,n) - y(m,n)| + \sum_{\substack{m,n \\ |x(m,n)-y(m,n)| > c}} \frac{[x(m,n) - y(m,n)]^2 + c^2}{2c} \quad (9)$$

$x_j, y_j$ are $2^{j-1}$ down-sampled images of $x, y$, respectively, $\mu_x, \sigma_x^2$ denote the mean and variance of $x$, respectively, and $\sigma_{xy}^2$ denotes the covariance between $x$ and $y$. $x(m, n)$ is the intensity value at pixel $(m, n)$ of image $x$. $\alpha_M, \beta_j, \gamma_j, C_i$ are empirical constants [64] and $c$ is a constant set as 0.1.

The loss for the discriminator $L_D$ is defined as:

$$L_D = \frac{1}{2}D(\hat{y})^2 + \frac{1}{2}[D(y) - 1]^2 \quad (10)$$

where $D$ is the discriminator of the GAN framework. An Adam optimizer [66] with an initial learning rate $10^{-5}$ was employed for stochastic optimization.

The training time on a PC with Intel Xeon W-2195 CPU, 256 GB RAM and one single NVIDIA RTX 2080 Ti graphic card is about 3 days. The reconstruction time using Recurrent-MZ ($M = 3$) of a volume of $101 \times 256 \times 256$ pixels takes ~2.2 s, and the reconstruction of an output image of $1024 \times 1024$ takes ~0.28 s.

### 4.4 The implementation of Deep-Z

The Deep-Z network, used for comparison purposes, is identical as in Ref. [38], and was trained and tested on the same dataset as Recurrent-MZ using the same machine. The loss function, optimizer and hyperparameter settings were also identical to Ref. [38]. Due to the single-scan propagation of Deep-Z, the training range is $\frac{1}{M}$ of that of Recurrent-MZ, depending on the value of $M$ used in the comparison. The reconstructed volumes over a large axial range, as presented in the manuscript, were axially stacked using $M$ non-overlapping volumes, which were propagated from different input scans and covered $\frac{1}{M}$ of the total axial range. The Deep-Z reconstruction time for a $1024 \times 1024$ output image on the same machine as Recurrent-MZ is ~0.12 s.

### 4.5 The implementation of 3D U-Net

For each input sequence of $M \times 256 \times 256 \times 2$ (the second channel is the DPM), it was reshaped as a tensor of $256 \times 256 \times (2M)$ and fed into the 3D U-Net. [46] When permuting the $M$ input scans, the DPMs always follow the corresponding images/scans. The number of channels at the last convolutional layer of each down-sampling block is $60 \cdot 2^k$ and the convolutional kernel is $3 \times 3 \times 3$. The network structure is the same as reported in Ref. [46]. The other training settings, such as the loss function and optimizer are similar to Recurrent-MZ. The reconstruction time ($M = 3$) for an output image of $1024 \times 1024$ on the same machine (Intel Xeon W-2195 CPU, 256 GB RAM and one single NVIDIA RTX 2080 Ti graphic card) is ~0.2 s, i.e., very similar to Recurrent-MZ inference time (0.28 sec) for the same inputs.